\documentclass[%
 reprint,
superscriptaddress,
 amsmath,amssymb,
 aps,
prx,
]{revtex4-1}

\usepackage{graphicx}
\usepackage{dcolumn}
\usepackage{bm}
\usepackage{braket}
\usepackage{amsthm}
\usepackage{siunitx}
\usepackage{gensymb}

\usepackage{hyperref}

\begin{document}

\newcommand{\ARCaffil}{\affiliation{IBM Quantum, Almaden Research Center, San Jose, California 95120, USA}}
\newcommand{\YKTaffil}{\affiliation{IBM Quantum, T. J. Watson Research Center, Yorktown Heights, NY 10598, USA}}
\newcommand{\ra}{\rangle}
\newcommand{\la}{\langle}
\newcommand{\be}{\begin{equation}}
\newcommand{\ee}{\end{equation}}
\newcommand{\EE}{\mathbb{E}}

\newcommand{\tO}{\tilde{O}}
\newcommand{\comment}[1]{(**)}
\newcommand{\onenorm}[1]{\left\Vert{#1}\right\Vert_1}

\newcommand{\ketbra}[2]{\ket{#1}\!\bra{#2}}

\newtheorem{dfn}{Definition}
\newtheorem{lemma}{Lemma}


\title{Doubling the size of quantum simulators by entanglement forging}

\author{Andrew Eddins}
\ARCaffil
\author{Mario Motta}
\ARCaffil
\author{Tanvi P. Gujarati}
\ARCaffil
\author{Sergey Bravyi}
\YKTaffil
\author{Antonio Mezzacapo}
\YKTaffil
\author{Charles Hadfield}
\YKTaffil
\author{Sarah Sheldon}
\ARCaffil

\date{\today}

\begin{abstract}
Quantum computers are promising for simulations of chemical and physical systems, but the limited capabilities of today's quantum processors permit only small, and often approximate, simulations. Here we present a method, classical entanglement forging, that harnesses classical resources to capture quantum correlations and double the size of the system that can be simulated on quantum hardware. Shifting some of the computation to classical post-processing allows us to represent ten spin-orbitals on five qubits of an IBM Quantum processor to compute the ground state energy of the water molecule in the most accurate simulation to date. We discuss conditions for applicability of classical entanglement forging and present a roadmap for scaling to larger problems.
\end{abstract}

\maketitle

Simulating quantum systems is an especially hard task for classical computers, making the realization of quantum computers potentially revolutionary for the study of chemistry, materials science, and fundamental physics. However, techniques like quantum phase estimation, which promises accurate chemical simulations, require hardware well beyond the present state of the art. While hardware capabilities continue to steadily advance, limitations on both quantity and quality of qubits are giving rise to a new family of algorithms that leverage additional classical resources to enable quantum computations requiring more qubits than physically available \cite{bravyi2016trading,peng2020simulating,yamazaki2018towards,kreula2016few,rubin2016hybrid, bauer2016hybrid, bravyi2017complexity,yuan2020quantum,mitarai2020constructing,kawashima2021efficient,Tang_2021}. Such algorithms broadly benefit from a strategy of partitioning a problem into weakly interacting clusters, then correlating the results of each on a classical computer. A range of important systems naturally possess suitable partitions, including low-energy eigenstates of chemical \cite{mcardle2020quantum} and lattice-model \cite{eisert2010colloquium,schollwock2011density,liu2019variational} Hamiltonians, systems embedded in a quantum bath \cite{kotliar2006electronic,sun2016quantum}, and static correlations associated with chemical bond-breaking processes \cite{bytautas2011seniority,elfving2020simulating,gunst2020seniority}. Such quantum-classical programming requires a profound understanding of the tradeoff between quantum and classical computational resources \cite{chitambar2019quantum} but is of theoretical and practical importance as it can extend the possibilities of classical and quantum computers alike.

\begin{figure*}
    \centering
    \includegraphics[width=0.7\linewidth]{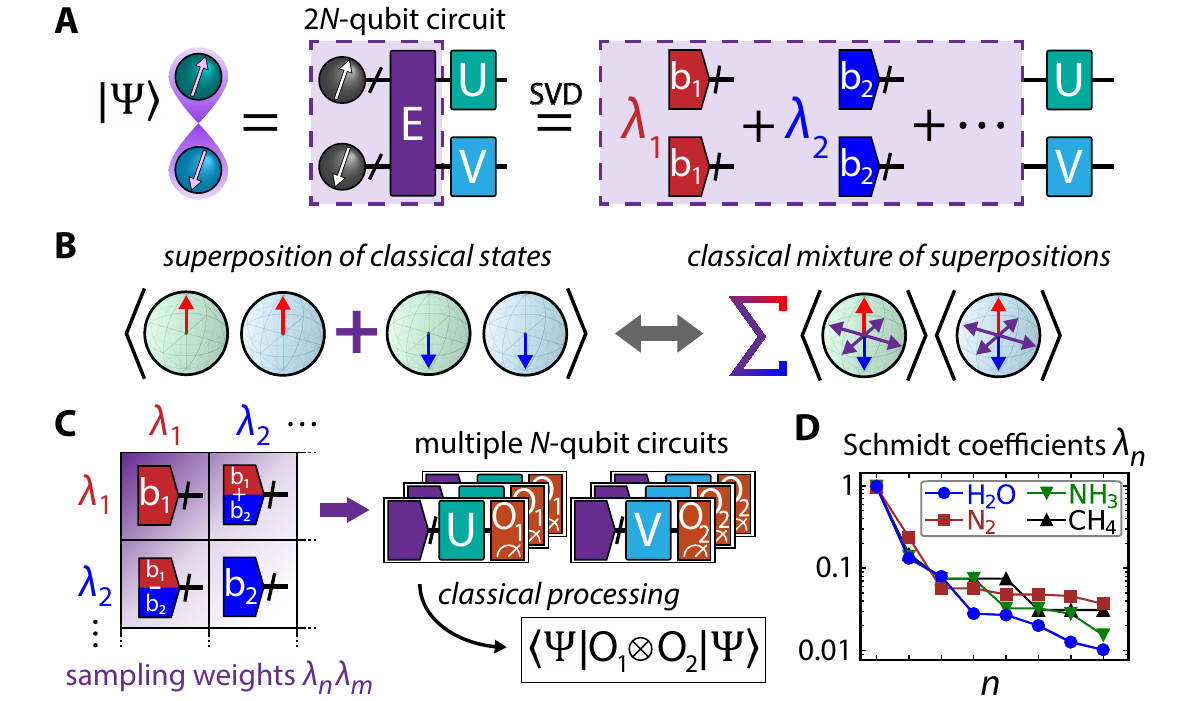}
    \caption{\textbf{Schematic overview of the entanglement forging protocol.} \textbf{A}, A state $\ket{\Psi}$ of a bipartite quantum system, here labeled with arrows alluding to spin polarization, can be defined by gates $E$, $U$, and $V$, where $E$ outputs a combination of bitstring states $\ket{b_n}\ket{b_n}$. \textbf{B}, A two-qubit entangled state can be rewritten using one-qubit superposition states. Changing labels $0,1\rightarrow b_n,b_m$ gives a transformation acting on components of the $2N$-qubit state. \textbf{C}, $\ket{\Psi}$ can be reconstructed from $N$-qubit circuits initialized as bitstrings and pairwise superpositions thereof. Circuits associated with small $\lambda_n\lambda_m$ can be estimated adequately from few samples. \textbf{D}, Rapid (slow) decay of the leading Schmidt coefficients in the decomposition of a molecular ground state signals weak (strong) entanglement between spin-up and spin-down particles.}
    \label{fig:EF_schematic}
\end{figure*}

Here we introduce a scheme we call \textit{classically forged entanglement}, which represents a $2N$-qubit wavefunction as multiple $N$-qubit states embedded in a classical computation. Beyond the reduction in requisite qubit number, offloading entanglement synthesis to classical processing permits the constituent $N$-qubit quantum circuits to be shallower, relaxing requirements on gate error and connectivity, at the cost of an increased number of circuit executions. In this report, we first theoretically describe entanglement forging, including scalable application to any state that can be partitioned into two weakly entangled halves. We then demonstrate entanglement forging using the variational quantum eigensolver (VQE) algorithm \cite{Peruzzo2014} to simulate a model of the water molecule \cite{nam2020ground}, representing 10 spin-orbitals with only five qubits. Conducted entirely via the IBM Quantum cloud-computing service, the experiment benefits from state-of-the-art capabilities \cite{jurcevic2020demonstration} such as active qubit reset, zero-noise extrapolation \cite{Kandala2019}, chip-level parallel execution, and a two-qubit ``hop'' gate tailored for chemical simulation.

We begin with Schmidt decomposition (Fig. \ref{fig:EF_schematic}A), a standard application of singular value decomposition (SVD) that allows one to write any state $\ket{\psi}$ of a bipartite $N+N$ qubit system as
\be
\ket{\psi} = (U\otimes V)\sum_{n=1}^{2^N} \lambda_n\ket{b_n} \otimes \ket{b_n}.
\ee
Here ${\ket{b_n}}$ are the $N$-qubit bitstring states, also known as computational basis states, $U, V$ are unitary operators respective to the two subsystems, and the Schmidt coefficients $\lambda_n$ may be taken to be non-negative. The flatter the distribution of Schmidt coefficients, the stronger the entanglement; a uniform distribution $\lambda_n=1/\sqrt{2^N}$ indicates the two halves of the system are maximally entangled, while only one nonzero coefficient corresponds to no entanglement.

Mixing classical and quantum information, entanglement-forging is more naturally expressed using density operators rather than wavefunctions, even for pure states such as $\ket{\psi}$. As shown in the supplementary material (\ref{sec:basicForging}), one can write the density operator as
\begin{align} \label{densityOp}
\begin{split}
&\ketbra{\psi}{\psi} = (U\otimes V)\sum_{n=1}^{2^N}\Big(\lambda_n^2 \ketbra{b_n}{b_n}^{\otimes 2} \\
&+\sum_{m=1}^{n-1} \lambda_n\lambda_m
\sum_{p\in \mathbb{Z}_4}(-1)^p \ketbra{\phi^p_{b_{n}b_{m}}}{\phi^p_{b_{n}b_{m}}}^{\otimes 2}
\Big)(U^\dag\otimes V^\dag),
\end{split}
\end{align}
where we have used the definition $\ket{\phi^p_{xy}} = \big(\ket{x}+i^p \ket{y}\big)/\sqrt{2}$ with $p\in \{0,1,2,3\}=\mathbb{Z}_4$. For example, in the minimal case of two qubits (Fig. \ref{fig:EF_schematic}B), $\ket{\phi^p_{01}}$ correspond to four equatorial points on Bloch sphere, rewriting a quantum superposition of product states in terms of classical products of superposition states. Eq. \eqref{densityOp} generalizes methods proposed in \cite{garcia2014geometry, PhysRevLett.122.230401, bravyi2016trading}, and is connected to tensor network representations of quantum circuits \cite{markov2008simulating,yuan2020quantum}, variational simulation of open quantum systems \cite{endo2020variational}, and the encoding of open-shell singlet and triplet states \cite{greene2021generalized}. The expectation of a $2N$-qubit operator $O = O_1 \otimes O_2$ is now
\begin{align} \label{expectation}
\begin{split}
\la O \ra = &\sum_{n=1}^{2^N}\Big(\lambda_n^2 \braket{b_n|\tO_1|b_n}\braket{b_n|\tO_2|b_n} +\sum_{m=1}^{n-1} \lambda_n\lambda_m \\
&\sum_{p\in \mathbb{Z}_4}(-1)^p \braket{\phi^p_{b_{n}b_{m}}|\tO_1|\phi^p_{b_{n}b_{m}}}\braket{\phi^p_{b_{n}b_{m}}|\tO_2|\phi^p_{b_{n}b_{m}}}\Big),
\end{split}
\end{align}
where $\tO_1 = U^\dag O_1 U$ and  $\tO_2 = V^\dag O_2V$, and each constituent requires only $N$ qubits to evaluate.

The resulting summation for $\la O \ra$ is not obviously scalable, involving as many as $2^{N+1}(2^{N+1}-1)$ distinct $N$-qubit quantum circuits. Nonetheless, if one restricts to simulations of sufficiently weak entanglement, $\la O \ra$ can be efficiently estimated (\ref{weightedSamplingTheory}) by sampling each circuit in proportion to the associated coefficients $\lambda_n\lambda_m$ in \eqref{expectation}, with a total number of samples for target precision $\epsilon$ scaling as 
\be
S \sim \Big(\frac{1}{\epsilon}\sum_{n,m} |\lambda_n \lambda_m |\Big)^2 = 
\frac{\|\vec{\lambda}\|_1^4}{\epsilon^2}\;,\;
\| \vec{\lambda} \|_1 = \sum_n |\lambda_n| \;.
\ee
Executing a quantum circuit once provides one sample of the corresponding expectation value, such that total runtime scales linearly with $S$. Since the one-norm decreases toward 1 in the limit of weak entanglement, the overhead cost of entanglement forging is smaller for simulations of states divisible into weakly-entangled halves, such as the spin-up and spin-down components of some molecular ground states (Fig.~\ref{fig:EF_schematic}D), and scales efficiently when the one-norm is at most polynomial in the problem size. For example, in some statically correlated ground states, $S$ can be independent of the number of basis orbitals. Outside of the domain of scalability, entanglement forging still enables useful heuristic simulations beyond the standard capacity of given quantum hardware, which may be realized with precision by truncating the list of bitstring states retained in the Schmidt decomposition.

Alternatively, this overhead may be reduced to a constant factor independent of qubit number via a complementary scheme (\ref{sec:HeisenbergForging}) simulating quantum correlations between subsystems using those within a subsystem, rather than using classical correlations as above. This method can be seen as an application of forging in the Heisenberg picture, reinterpreting an observable acting on $N+N$ qubits as a classical mixture of operators describing the forward and backward time evolution of $N$-qubits, at a cost of deeper circuits. Provided certain sampling assumptions, this method is not limited to weakly entangled states, so it may be applicable to a wider range of systems.

\begin{figure*}
    \centering
    \includegraphics[width=0.7\linewidth]{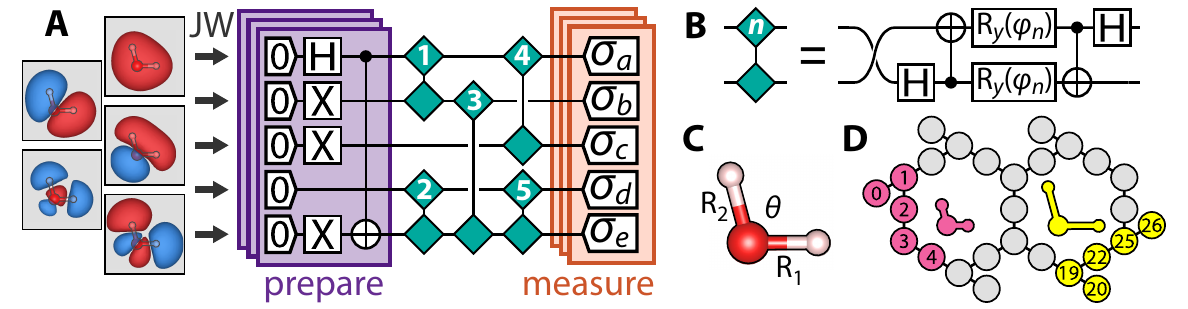}
    \caption{\textbf{Experimental ansatz structure.} \textbf{A}, The five active molecular orbitals are encoded in a line of qubits via the Jordan-Wigner mapping. The qubits are prepared in either a bitstring state, or a superposition of two bitstrings (shown), and then acted upon by parameterized hop gates (\textbf{B}). In order to evaluate the Pauli strings of the Hamiltonian, each state is prepared many times, and rotated appropriately prior to measurement in the computational basis. \textbf{C}, Parameterized molecular geometry from which molecular orbitals are determined. \textbf{D}, Device map of \textit{ibmq\_dublin}, with a highlighted example of how two geometries may be solved in parallel.}
    \label{fig:ansatz}
\end{figure*}

We use entanglement forging for a VQE simulation of the water molecule, as schematized in Figure \ref{fig:ansatz}. Starting from the minimal STO-6G basis, we freeze the core oxygen $1s$ and the out-of-plane oxygen $2p$ orbitals, leaving an active space of 10 spin-orbitals. The Jordan Wigner mapping encodes either the spin-up or spin-down orbitals onto each five-qubit quantum circuit, and we simplify the problem structure by asserting $V=U$ per the known symmetry between spin-polarizations in the closed-shell singlet ground state.

How best to construct an ansatz circuit for VQE remains an open and active research topic \cite{cerezo2020variational}. To facilitate our demonstration of entanglement forging, we first ran classical simulations of VQE without entanglement forging to obtain a 10-qubit circuit that performs well for this problem. Based on this circuit, we selected the quantum gates in $U$, and truncated the Schmidt decomposition to $k=3$ of the 10 possible bitstrings, namely
$\ket{b_1} = \ket{11100}, \ket{b_2} = \ket{01110},$ and $\ket{b_3} = \ket{01101}$. Such truncation was convenient due to a technical limitation on the number of unique circuits, distinct from the more fundamental limitation of overhead time scaling with the number of samples $S$ discussed above; future implementations may variationally explore larger subsets of possible bitstrings as discussed in \ref{weightedSamplingTheory}. The resulting ansatz is thus tailored to our particular problem and parameterized both by two qubit gate rotation angles and by the Schmidt coefficients, discussed further below. To realize each circuit required for equation \eqref{expectation}, we prepare either one bitstring, or a superposition of two. For example, Fig. \ref{fig:ansatz}A shows how to prepare $\ket{\phi^0_{b_1b_3}}$, and a general construction is given in \ref{stateprep}. 

Each initialized state is then acted upon by the unitary $U$, here comprised of five two-qubit ``hop gates'' (Fig. \ref{fig:ansatz}A,B), where each acts according to the matrix
\begin{equation}
h(\varphi) =
\begin{bmatrix}
1 & 0 & 0 & 0\\
0 & \cos(\varphi) & -\sin(\varphi) & 0\\
0 & \sin(\varphi) & \cos(\varphi) & 0\\
0 & 0 & 0 & -1
\end{bmatrix}.
\end{equation}
In the simulated molecule, the rotation by $\varphi$ moves particles between orbitals, while the entry -1 provides a CPHASE-like two-particle interaction. A product of hop gates is universal with respect to real-valued wavefunctions of fixed particle number, making the gate appealing for simulating eigenstates of time-reversal invariant Hamiltonians. Details of compilation on quantum hardware appear in \ref{connectivity}.

The runtime cost of entanglement forging makes fast circuit execution critical. We implemented the VQE routine using the Qiskit quantum computing package, and executed the experiment via the IBM Quantum cloud computing service on the \textit{ibmq\_dublin} processor \cite{qiskit, dublin}. Active qubit reset enabled execution at a 10 kHz repetition rate. Noting that distinct molecular geometries present embarrassingly parallel problems, we mapped two independent problems onto separate five-qubit chains (Fig. \ref{fig:ansatz}D), doubling throughput to 20,000 five-qubit circuits per second  (\ref{parallelization}). While freedom to independently specify the number of samples for each unique circuit would best economize runtime, here we approximated this behavior by submitting multiple copies of circuits in proportion to the desired weighting (\ref{weightedSamplingExpt}).

\begin{figure*}
    \centering
    \includegraphics[width=0.7\linewidth]{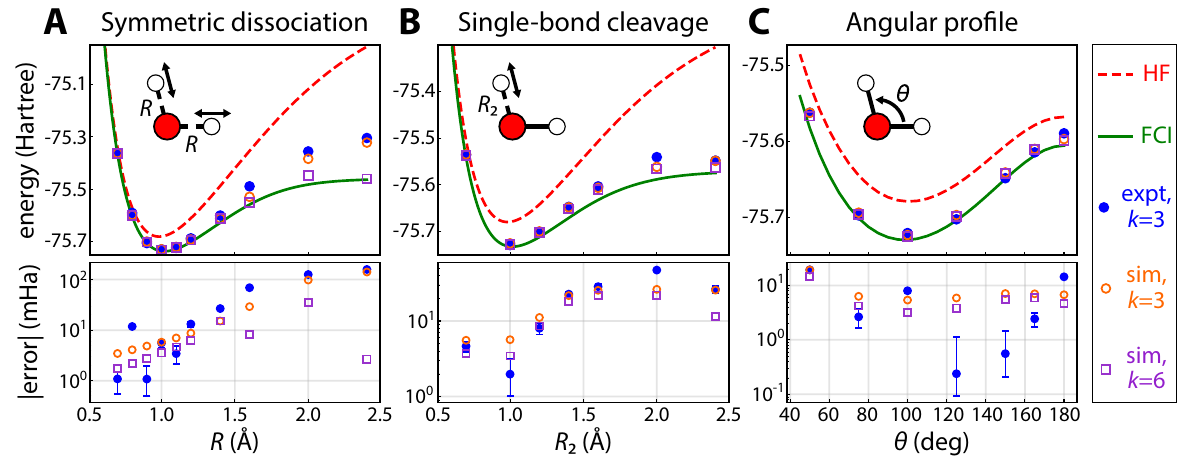}
    \caption{\textbf{VQE energies.} Ground-state energies computed while varying \textbf{A}, both O-H bond lengths, \textbf{B}, a single O-H bond length, and \textbf{C}, the H-O-H angle $\theta$. In the upper plots, VQE results using the entanglement-forging ansatz on \textit{ibmq\_dublin} appear as filled blue circles, alongside curves indicating the classically computed Hartree Fock (HF) and full configuration interaction (FCI) values for the active space. Unfilled shapes indicate results from running VQE on a noiseless classical simulator using the same 3-bitstring ansatz used on \textit{ibmq\_dublin} ($k=3$), and using a larger 6-bitstring ansatz ($k=6$). Lower plots indicate absolute errors, $|E-E_{\mathrm{FCI}}|$. Error bars are one standard deviation produced by bootstrapping measurement histograms, representing precision but not systematic error or drift.}
    \label{fig:energy_plots}
\end{figure*}

\begin{figure*}
    \centering
    \includegraphics[width=0.7\linewidth]{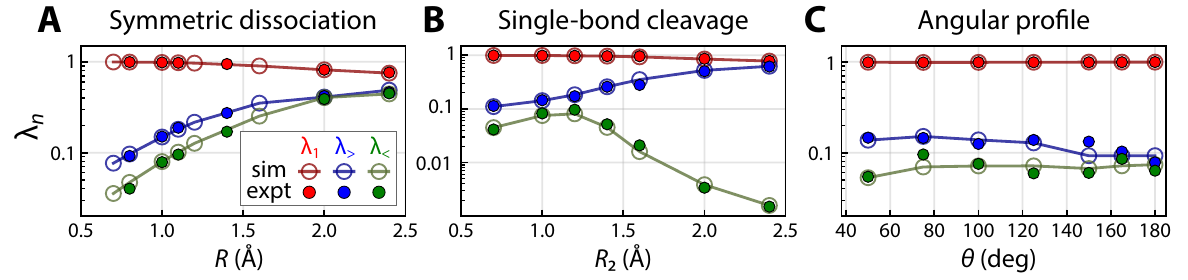}
    \caption{\textbf{Entanglement structure.} Schmidt coefficients $\lambda_n$ obtained from VQE run on \textit{ibmq\_dublin} (solid circles) and on a noiseless classical simulator (empty circles) while varying molecular geometry as in Fig. \ref{fig:energy_plots}. The largest component is the Hartree Fock amplitude ($\lambda_1$, red); the two remaining values are sorted and plotted in blue and green. \textbf{A}, Bitstrings beyond Hartree Fock become increasingly significant as the molecular bonds are stretched. \textbf{B}, Removing a single hydrogen produces a ground state wavefunction dominated by two bitstrings. \textbf{C}, All choices of H-O-H bond angle yielded ground states characterized by weak entanglement.} 
    \label{fig:singular_values}
\end{figure*}

Properties of the particular problem and ansatz permitted many circuits to be omitted, further accelerating execution. Generally, superposition states $\ket{\phi_{b_nb_m}^p}$ may be skipped when evaluating Hamiltonian terms acting on only one of the two subsystems. Likewise, since here the operators $O_1$ and $O_2$ in the Hamiltonian are real valued in the bitstring basis, contributions from $\ket{\phi_{b_nb_m}^p}$ with odd $p$ must vanish. Finally, observing that our chosen hop gates nominally do not modify the Hartree-Fock (HF) state, we omit the corresponding circuits and use instead the classical mean-field result, fixing $\braket{b_1|\tO|b_1} = E_{\mathrm{HF}}$. This simplification cannot capture excitations of unpaired electrons across the HF occupied-virtual threshold, and risks unphysical results as hardware errors impact intermediate calculations unevenly. Nonetheless, unburdening the quantum processor of the dominant, classically-accessible part of the problem reduced $S$ by roughly an order of magnitude for our near-equilibrium experiments, while retaining acceptable overall accuracy. Notably, this acceleration was enabled by the mixture of classical and quantum elements in the entanglement-forging representation.

VQE was repeated to study the behavior of entanglement forging across a variety of molecular geometries. For each geometry, optimization was performed in approximately 100 iterations of an SPSA (simultaneous perturbation stochastic approximation) algorithm \cite{spall1998overview,adaptSPSAIto2016}. Figure \ref{fig:energy_plots} displays the energies resulting from varying the length of both O-H bonds, the length of only one bond, and the H-O-H angle at fixed bond length. Where not otherwise specified, nominal equilibrium values of $R_{\mathrm{eq}}=\SI{0.958}{\angstrom}$ and $\theta_{\mathrm{eq}}=104.478\degree$ were used \cite{cccbdb}. Deviations between the VQE results and exact full configuration interaction (FCI) energies trend from $\sim1-10$ mHa to $\sim10-100$ mHa as bond lengths are increased, in line with the expectation that entanglement forging should work best for problems with weakly entangled ground states. We repeated all experiments on a noiseless classical simulator, first verifying that this trend persists in the absence of gate errors (orange circles), and second observing that accuracy in the stretched regime improves substantially upon increasing the number of represented bitstring states $k$ from three to six (purple squares), enabling the forging to describe stronger entanglement at the expense of sampling more distinct circuits. Finally, to emphasize that entanglement forging can extend to arbitrary accuracy with no increase in the number of qubits, we ran a noiseless VQE simulation at the nominal equilibrium geometry including all 10 three-occupation bitstrings along with more hop gates, which converged within 1.6 mHa of the FCI energy (\ref{largerAnsatz}). 

The Schmidt coefficients $\lambda_n$ (Fig. \ref{fig:EF_schematic}B) are among the parameters optimized by VQE with entanglement forging (\ref{updatingLambda}). These coefficients are uniquely defined by the ground-state wavefunction and choice of partition, and thus, assuming adequate convergence of VQE to the ground state, do not depend on the selection of gates in $U$ and $V$ up to reordering of the index $n$. Figure \ref{fig:singular_values} shows the Schmidt coefficients obtained on \textit{ibmq\_dublin}, along with those obtained on a noiseless classical simulator with the same 3-bitstring ansatz ($k=3$). The plots verify that weaker entanglement correlates with better performance in Fig. \ref{fig:energy_plots}, and clarify how bond stretching strengthens spin-polarization entanglement. For instance, removing a single hydrogen (Fig. \ref{fig:singular_values}B) leads to a wavefunction dominated by two bitstrings, consistent with a picture of dissociation into spin-1/2 fragments with opposite polarizations. Further discussion of Schmidt coefficient distributions supported by FCI calculations is provided in \ref{svdTruncation}.

We have demonstrated entanglement forging by computing the ground state energy of a water molecule for varying geometries using five qubits to represent ten spin-orbitals, the most accurate VQE simulation of this molecule using quantum hardware to date. Entanglement forging opens a vista of experimental possibilities, such as dividing a lattice system using a spatial- rather than spin-partition; introducing an adaptive bisection based on orbital optimization; using entanglement forging in simulation algorithms beyond VQE, including error-corrected techniques such as quantum phase estimation; and more generally for any experiment otherwise inaccessible due to limited qubit number or connectivity where the overhead of forging may be tolerated. 
Where overhead remains prohibitive, Heisenberg-picture forging (\ref{sec:HeisenbergForging}) may provide an alternative path to simulation of large systems. Additional exploration of techniques incorporating classical resources with quantum processing may further enhance the power of near term quantum simulations and help realize quantum advantage for practical applications.

\bibliography{bibliograph}

\bibliographystyle{sty}

\textbf{Acknowledgments:} We acknowledge Doug McClure, Thomas Alexander, Stephen Wood, Youngseok Kim, Daniel Egger, and the IBM Quantum backend team for technical assistance, and Julia Rice, John Lapeyre, Agata Branczyk, Lev Bishop, and Jay Gambetta for valuable discussions. \textbf{Funding:} SB is supported in part by the Army Research Office under Grant Number W911NF-20-1-0014 and by the IBM Research Frontiers Institute. \textbf{Authors contributions:} SB conceptualized the project. AE, MM, TG, SS developed the main experiment and analysis software, and ran simulations and measurements. SB, AM, CH provided theoretical modeling and analyses. All authors contributed to the manuscript. \textbf{Competing interests:} Elements of this work are included in a patent filed by the International Business Machines Corporation with the US Patent and Trademark Office. \textbf{Data and materials availability:} Experiment data and code available upon reasonable request to the authors.

\clearpage

\onecolumngrid

\renewcommand{\thesubsection}{SM.\arabic{subsection}}

\begin{center}
\section*{Supplemental material}
\end{center}

\subsection{Decomposition to superpositions of $N$-qubit bitstrings}
\label{sec:basicForging}
We would like to forge a 2$N$-qubit state
\[
\ket{\Psi} = \sum_{n}\lambda_n \ket{b_n}\otimes\ket{b_n}
\]
where $\lambda_n$ are real coefficients. For each distinct pair of bitstrings $x,y$ and each $p\in \{0,1,2,3\} \equiv \mathbb{Z}_4$ define a state
\[
\ket{\phi_{xy}^p} = \frac{\ket{x}+i^p\ket{y}}{\sqrt{2}}.
\]
One gets
\[
4\ketbra{\phi_{xy}^p}{\phi_{xy}^p}^{\otimes 2} = i^{2p}\ketbra{y}{x}^{\otimes 2} + i^{-2p}\ketbra{x}{y}^{\otimes 2} + O + i^p O' + i^{-p}O''
\]
where $O,O',O''$ are some operators. Note that $\sum_{p\in\mathbb{Z}_4}i^{pq} = 4\delta_{q,0}$ for all $q \in \mathbb{Z}_4$. Thus
\[
\sum_{p\in\mathbb{Z}_4}i^{2p}\ketbra{\phi_{xy}^p}{\phi_{xy}^p}^{\otimes 2} = \ketbra{y}{x}^{\otimes 2} + \ketbra{x}{y}^{\otimes 2}.
\]
Applying the above identity to each off-diagonal term in $\ketbra{\Psi}{\Psi}$ we arrive at
\[
\ketbra{\Psi}{\Psi} = \sum_n \lambda_n^2\ketbra{b_n}{b_n}^{\otimes 2} + \sum_n\sum_{m<n}\lambda_n \lambda_m \sum_{p\in \mathbb{Z}_4}(-1)^p\ketbra{\phi_{b_n b_m}^p}{\phi_{b_n b_m}^p}^{\otimes 2} .
\]

\subsection{Entanglement forging in the Heisenberg picture}
\label{sec:HeisenbergForging}

First let us define a class of quantum states whose entanglement can be efficiently forged.
Consider a system of $2N$ qubits and  states of the form
\begin{equation}
\label{scalable_ansatz}
|\psi\rangle =(U\otimes U) \sum_{x\in \{0,1\}^N}  \lambda_x  |x\rangle \otimes |x\rangle,
\end{equation}
where the tensor product separates two $N$-qubit registers, $\lambda_x$ are real coefficients,
and $U$ is a unitary operator with real matrix elements in the standard basis.
The coefficients $\lambda_x$ have to be normalized such that $\sum_x \lambda_x^2=1$.
We assume that $U$ admits an efficient implementation by a quantum circuit.
It is natural to consider quantum circuits composed of 
real gates such as the Hadamard, $Y$-rotation, CNOT, or a hop gate.
This ensures that $U$ has real matrix elements.
The vector of $2^N$ coefficients $\lambda_x$
may be specified either explicitly (for small $N$) or implicitly
by  a classical algorithm that 
can efficiently perform the following tasks:\\
(i) sample a bit string $x$ from the probability distribution $\lambda_x^2$;\\
(ii) compute the ratio $\lambda_y/\lambda_x$ for a given pair of bit strings
$x,y$.\\
Such implicit description of exponentially large vectors
is commonly used in Quantum Monte Carlo simulations
based on tensor networks~\cite{sandvik2007variational} or neural network states~\cite{carleo2017solving}.
It follows the framework of computationally tractable quantum states introduced in~\cite{nest2009simulating}.

Given Pauli observables $O_1,O_2\in \{I,X,Y,Z\}^{\otimes N}$,
our goal is to estimate the expected value 
\begin{equation}
\label{eq2}
\mu = \langle \psi | O_1 \otimes O_2 |\psi\rangle.
\end{equation}
We show how to accomplish this task by a series of
$N$-qubit experiments.
A typical experiment 
prepares a state  $U^\dag C U|x\rangle$ 
for a suitable $N$-qubit Clifford circuit $C$ and a bit string $x$.
Then every qubit of the state $U^\dag C U|x\rangle$  is measured in the standard basis.
Note that the circuit $U^\dag$ can be obtained from $U$ by inverting the order of gates
and replacing each gate by its inverse.
The size of the Clifford circuit $C$ depends on the form of the observables $O_1,O_2$.
Let $|O_j|$ be the Hamming weight of $O_j$, that is, the number of 
single-qubit terms $X$, $Y$, $Z$ that appear in $O_j$. 
Then $C$ contains at most  $2(|O_1|+|O_2|)$ CNOT gates and some single-qubit
gates. 
The number of experiments required to approximate
the expected value Eq.~(\ref{eq2})  with a precision $\epsilon$
is proportional to $1/\epsilon^2$ with a constant prefactor. 
Crucially, the number of experiments  does not depend on the number of qubits $N$
or the amount of entanglement in the forged state $|\psi\rangle$.

The key idea behind this method is to convert
a Pauli observable $O_1 \otimes  O_2 $ describing a bipartite
system of $N+N$ qubits into a classical mixture of observables
$C^* \otimes C$, where $C$ is a self-adjoint $N$-qubit Clifford operator
and $C^*$ is the complex conjugate of $C$. 
By interpreting $C$ and $C^*$  as operators
describing the forward and the backward time evolution of the same $N$-qubit register
we express the expected value $\mu$ as a mixture of
quantum probabilities that can be measured on an $N$-qubit device.
Since the analysis of the method is performed  in the
Heisenberg picture at the level of observables, we shall refer to it as the {\em Heisenberg
forging}. Meanwhile the forging method described in the main text works primarily in the
Schr\"odinger picture by decomposing an entangled state of $N+N$ qubits into
a classical mixture of product states. We shall refer to the latter as 
the {\em Schr\"odinger forging}, to avoid a confusion between the two methods.

The Heisenberg forging roughly doubles the size
of quantum circuits that need to be executed compared with the Schr\"odinger forging
(the former needs to implement both $U$ and $U^\dag$ while the latter
only needs to implement $U$).  However, the Heisenberg  forging requires shorter circuits compared with
the   preparation of the full state $|\psi\rangle$
on a hardware with $2N$ qubits. Indeed, a quantum circuit preparing the state $|\psi\rangle$
includes two copies of $U$ as well as some extra gates generating
the initial entanglement between the $N$-qubit registers.
Thus we expect that the Heisenberg forging reduces both the number of qubits and the number
of gates per  experiment while increasing the number of experiments
only by a constant factor.

Using states Eq.~(\ref{scalable_ansatz}) as a variational ansatz in quantum
simulations is justified if the exact ground state has real amplitudes in the standard
basis and is invariant under exchanging the two $N$-qubit registers.
As discussed in the main text, this is the case for molecular electronic structure Hamiltonians
assuming that the $N$-qubit registers represent spin-up and spin-down orbitals.
Specializing Eq.~(\ref{scalable_ansatz}) to the case $U=I$ gives
variational states in which every spatial orbital is either empty or occupied by 
a pair of electrons in the singlet state.
Such states play the central role in the Restricted Hartree-Fock space formalism~\cite{helgaker2014molecular}.
As shown in Ref.~\cite{elfving2020simulating},
minimizing the energy of a molecular Hamiltonian over variational states  
of the form
Eq.~(\ref{scalable_ansatz}) with $U=I$
is equivalent to minimizing the energy of a quantum spin Hamiltonian
with two-spin Heisenberg-type interactions and an external magnetic field 
over states $|\lambda\rangle = \sum_x \lambda_x |x\rangle$,
where $\lambda_x$ are the Schmidt coefficients that appear in Eq.~(\ref{scalable_ansatz}).
We anticipate that the ansatz Eq.~(\ref{scalable_ansatz})  with $U\ne I$ may 
improve upon the Restricted Hartree-Fock formulation since it is based on
more general variational states. 
We leave applications of the Heisenberg forging for a future work. 

In the rest of this section  we assume that $O_1$  and $O_2$ are non-identity Pauli operators,
which is the most interesting case. 
Otherwise we have $O_1=I$ or $O_2=I$. For concreteness,
assume $O_2=I$. Then the expected value Eq.~(\ref{eq2}) becomes
$\mu =\sum_{x} \lambda_x^2 \langle x|U^\dag O_1 U|x\rangle$.
One can estimate $\mu$ with a precision $\epsilon$ by
generating $M\sim \epsilon^{-2}$ samples $x^1,\ldots,x^M\in \{0,1\}^N$ from the distribution $\lambda_x^2$,
measuring the eigenvalue of $O_1$ on each state $U|x^j\rangle$, and computing
the sample mean of the measured eigenvalues.

Let $\{O_1,O_2\} = O_1 O_2 + O_2 O_1$.
Below we construct a  decomposition 
\begin{equation}
\label{HF1}
O_1 \otimes O_2 + O_2 \otimes O_1 =
\frac{a_0}2\left(  \{O_1, O_2\} \otimes I  + I \otimes  \{O_1, O_2\}  \right)
+ 
 \sum_{j=1}^4 a_j C_j^* \otimes C_j,
\end{equation}
where $a_j$ are real coefficients such that $|a_j|\le 1$,
and $C_j$ are $N$-qubit Clifford operators.
Recall that the state $|\psi\rangle$ is invariant under the swap of the two $N$-qubit registers.
Thus  the expected value $\mu$ is invariant under swapping the observables $O_1$ and $O_2$.
Symmetrizing $\mu$ over the swap of $O_1$ and $O_2$ one gets 
$\mu = \left( \langle \psi |O_1 \otimes O_2 |\psi\rangle +  \langle \psi |O_2 \otimes O_1 |\psi\rangle \right)/2$.
Combining this and 
 Eq.~(\ref{HF1}) gives
\begin{equation}
\label{HF2}
\mu = a_0 \sum_x \lambda_x^2\,  \mathrm{Re}(\langle x| U^\dag O_1 O_2 U|x\rangle)  +  \frac12 \sum_{j=1}^4 a_j \langle \psi | C_j ^*\otimes C_j|\psi\rangle.
\end{equation}
We shall now convert the expected value
$\langle \psi | C_j^* \otimes C_j|\psi\rangle$ into a mixture of quantum probabilities that can be
measured on an $N$-qubit device.  From Eq.~(\ref{scalable_ansatz}) one gets
\begin{equation}
\label{HF3}
\langle \psi |C_j^*\otimes C_j|\psi\rangle =
\sum_{x,y} \lambda_x \lambda_y \langle y|U^\dag C_j^* U |x\rangle \langle y|U^\dag C_j U |x\rangle
=
\sum_{x,y} \lambda_x \lambda_y |\langle y | U^\dag C_j  U|x\rangle |^2.
\end{equation}
Here we used the identity
\[
\langle y|U^\dag C_j^* U |x\rangle = \langle x|(U^\dag C_j^* U)^T |y\rangle
= \langle x| U^T C_j^\dag U^*|y\rangle = \langle x|U^\dag C_j^\dag U|y\rangle
\]
which follows from the assumption that  $U$ has  real matrix elements.
Combining Eqs.~(\ref{HF2},\ref{HF3}) one gets
\begin{equation}
\label{HF4}
\mu = a_0 \sum_x \lambda_x^2\,  \mathrm{Re}(\langle x| U^\dag O_1 O_2 U|x\rangle) 
 +  \frac12 \sum_{j=1}^4 a_j \mu_j,
\end{equation}
where
\begin{equation}
\label{HF5}
\mu_j =  \sum_{x,y} \lambda_x \lambda_y |\langle y | U^\dag C_j  U|x\rangle |^2.
\end{equation}
The first term in Eq.~(\ref{HF4})  can be estimated on an $N$-qubit device by 
sampling a bit string $x$ from the distribution $\lambda_x^2$,
preparing a state $U|x\rangle$, and measuring the eigenvalue of $O_1 O_2$
(note that the first term in Eq.~(\ref{HF4}) is non-zero only if $O_1 O_2$ is a self-adjoint operator,
that is, if $O_1$ commutes with $O_2$). 
Below we focus on the second term in Eq.~(\ref{HF4}) and show how to estimate 
the quantities $\mu_j$.
Define a function
\begin{equation}
\label{HF6}
R(x,y) =\frac{\lambda_y}{\lambda_x}.
\end{equation}
By assumption, one can efficiently compute $R(x,y)$ for a given pair $x,y$.
For each bit string $x$ define a conditional probability distribution 
\begin{equation}
\label{HF7}
P_j(y|x) =  |\langle y | U^\dag C_j  U|x\rangle |^2.
\end{equation}
One can sample a bit string $y$ from $P_j(y|x)$ on a quantum device with $N$ qubits 
by preparing the state $U^\dag C_j U|x\rangle$
and measuring every qubit in the standard basis. Furthermore,
Eq.~(\ref{HF5}) implies that
$\mu_j$ is the expected value of $R(x,y)$ over the probability distribution
$\lambda_x^2 P_j(y|x)$, that is, 
\begin{equation}
\mu_j = \sum_{x,y} \lambda_x^2 P_j(y|x) R(x,y) \equiv \EE(R).
\end{equation}
The random variable $R(x,y)$ has the variance at most one since
\begin{equation}
\label{HF8}
\EE (R^2) = \sum_{x,y} \lambda_x^2  P_j(y|x) R^2(x,y)
= \sum_{x,y} \lambda_y^2\,  |\langle y | U^\dag C_j  U|x\rangle |^2 = \sum_y \lambda_y^2 = 1.
\end{equation}
Thus one can estimate $\mu_j$ with an additive error $\epsilon$ by generating
roughly $1/\epsilon^2$ samples $(x,y)$ from the distribution $\lambda_x^2 P_j(y|x)$
and computing the sample mean of $R(x,y)$.  
By estimating each term $\mu_j$ in Eq.~(\ref{HF4}) 
with a precision $\epsilon/4$ 
and estimating the expected value
$\mathrm{Re}(\langle x| U^\dag O_1 O_2 U|x\rangle) $ with a precision $\epsilon/2$
gives the desired $\epsilon$-estimate of $\mu$
(use the triangle inequality and the assumption $|a_j|\le 1$).
 The total number of $N$-qubit experiments
required to obtain these estimates is proportional to $1/\epsilon^2$
with a constant prefactor.

In the rest of this section we explicitly  construct the decomposition Eq.~(\ref{HF1})
for non-identity Pauli observables $O_1,O_2\in \{I,X,Y,Z\}^{\otimes N}$.
Let $w_j$ be the number of single-qubit $Y$
terms that appear in $O_j$.
Note that $O_j^*=(-1)^{w_j} O_j$, that is, the matrix of $O_j$ in the standard basis is real if $w_j$ is even
and imaginary if $w_j$ is odd. 
Since the state $|\psi\rangle$ is real, one has $\mu =\langle \psi|O_1\otimes O_2|\psi\rangle=0$
unless $w_1$ and $w_2$ have the same parity.
Below we assume that this is the case, that is,
$O_j^*=\sigma O_j$ for $\sigma=(-1)^{w_1}=(-1)^{w_2}$.
Recall that Pauli observables either commute or anti-commute.

{\em Case~1:} $O_1$ anti-commutes with $O_2$. 
Below we prove the following simple fact.
\begin{lemma}
\label{lemma:1}
Let $O_1,O_2 \in \{I,X,Y,Z\}^{\otimes N}$ be anti-commuting Pauli observables.
There exist $N$-qubit Clifford circuit $V$ and a qubit $q\in \{1,2,\ldots,N\}$ such that 
\begin{equation}
\label{HF9}
O_1=VX_q V^\dag \quad \mbox{and} \quad O_2 = V Z_q V^\dag.
\end{equation}
The circuit $V$ contains at most  $|O_1|+|O_2|-2$ CNOT gates and some single-qubit
Clifford gates. Here $|O_j|$ is  the Hamming weight $O_j$.
\end{lemma}
Let $V$ be the Clifford circuit from Lemma~\ref{lemma:1}. 
Define Clifford operators
\begin{equation}
\label{HF10}
C_1 = V \frac1{\sqrt{2}} (X_q + Z_q) V^\dag
\quad \mbox{and} \quad 
C_2 = V  \frac1{\sqrt{2}} (X_q - Z_q) V^\dag.
\end{equation}
Note that $C_1=(O_1+O_2)/\sqrt{2}$
and $C_2=(O_1-O_2)/\sqrt{2}$.
A simple algebra gives
\begin{equation}
\label{HF11}
O_1 \otimes O_2 + O_2 \otimes O_1 =  C_1 \otimes C_1 - C_2 \otimes C_2.
\end{equation}
Taking the partial transpose of Eq.~(\ref{HF11}) over the first
$N$-qubit register, using the fact that $O_j$ and $C_j$ are self-adjoint,
and the identity $O_j^*=\sigma O_j$ one gets
\begin{equation}
\label{HF12}
O_1 \otimes O_2 + O_2 \otimes O_1 = \sigma \left( C_1^* \otimes C_1 - C_2^* \otimes C_2\right).
\end{equation}
This is the desired decomposition Eq.~(\ref{HF1}) with
$a_1=\sigma$, $a_2=-\sigma$, and $a_0=a_3=a_4=0$.
Lemma~\ref{lemma:1} and Eq.~(\ref{HF10}) imply that $C_1$
and $C_2$ can be implemented by circuits with at most $2(|O_1|+|O_2|)$ CNOTs.

{\em Case~2:} $O_1$ commutes with $O_2$.
If $O_1=O_2$ then  choose $C_1=C_2=O_1$,
$a_1=a_2=\sigma$, and $a_0=a_3=a_4=0$.
One can easily check that Eq.~(\ref{HF1}) is satisfied. 
From now on we assume $O_1\ne O_2$.
Below we prove the following lemma.
\begin{lemma}
\label{lemma:2}
Let $O_1,O_2 \in \{I,X,Y,Z\}^{\otimes N}$ be commuting  Pauli observables
such that $O_1\ne O_2\ne I$.
There exist $N$-qubit Clifford circuit $V$ 
and a pair of qubits $p,q\in \{1,2,\ldots,N\}$ 
such that 
\begin{equation}
\label{HF13}
O_1=V Z_p V^\dag \quad \mbox{and} \quad O_2 = V Z_q V^\dag.
\end{equation}
The circuit $V$ contains at most  $|O_1|+|O_2|-2$ CNOT gates and some single-qubit
Clifford gates. 
\end{lemma}
Let $V$ be the Clifford circuit from Lemma~\ref{lemma:2}. 
Define $N$-qubit operators
\[
C_{\alpha,\beta} = V X_p^{\alpha} X_q^{\beta} \mathsf{CZ}_{p,q}  X_p^{\alpha} X_q^{\beta} V^\dag
\]
where $\alpha,\beta\in \{0,1\}$ and $\mathsf{CZ}_{p,q}$ denotes the controlled-$Z$ gate
acting on qubits $p,q$. Using the identity 
\[
X_p^{\alpha} X_q^{\beta}\mathsf{CZ}_{p,q}X_p^{\alpha} X_q^{\beta}=
(1/2)(I + (-1)^\alpha Z_p + (-1)^\beta Z_q - (-1)^{\alpha+\beta} Z_p Z_q)
\]
one gets
\[
C_{\alpha,\beta} = (1/2)(I + (-1)^\alpha O_1 + (-1)^\beta O_2 - (-1)^{\alpha+\beta} O_1 O_2).
\]
A simple algebra shows that
\[
O_1 \otimes O_2 + O_2 \otimes O_1 = 
O_1 O_2 \otimes I + I\otimes O_1 O_2 + \sum_{\alpha,\beta=0,1} \;
(-1)^{\alpha+\beta} C_{\alpha,\beta} \otimes C_{\alpha,\beta}.
\]
Taking the partial transpose over the first
$N$-qubit register, using the 
identity $O_j^*=\sigma O_j$, and the fact that $C_j$
and $O_j$  are self-adjoint one gets
\[
O_1 \otimes O_2 + O_2 \otimes O_1 = 
\sigma(O_1 O_2 \otimes I + I\otimes O_1 O_2) + \sigma \sum_{\alpha,\beta=0,1} \;
(-1)^{\alpha+\beta} C_{\alpha,\beta}^* \otimes C_{\alpha,\beta}.
\]
This is the desired decomposition Eq.~(\ref{HF1}) with
$C_1=C_{0,0}$, $C_2=C_{0,1}$, $C_3=C_{1,0}$, $C_4=C_{1,1}$,
$a_0=a_1=a_4=\sigma$, and $a_2=a_3=-\sigma$.

It remains to prove Lemmas~\ref{lemma:1},\ref{lemma:2}.
\begin{proof}[\bf Proof of Lemma~\ref{lemma:1}]	
We shall convert $O_1$ and $O_2$ to single-qubit Pauli operators
$X_q$ and $Z_q$ respectively by a sequence of  steps
$O_1\gets W^\dag O_1 W$ and $O_2 \gets W^\dag O_2 W$,
where $W$ is a Clifford circuit composed of $\mathsf{CNOT}$, $\mathsf{CZ}$, and single-qubit gates. 
We shall choose $W$ such that at each step  the combined weight $|O_1|+|O_2|$
is reduced at least by number of two-qubit gates in $W$.
The desired circuit $V$ is then obtained as the composition of the
circuits $W$ applied at each step. 

The first step converts
 $O_1$ and $O_2$ into the standard form such that
their action on any qubit falls into one of  five cases shown below.
\begin{center}
	\begin{tabular}{|c|c|c|c|c|c|}
		\hline 
		Case & A & B & C & D & E \\
		\hline 
		\hline
		$O_1$ & $X$ & $Z$ & $I$ & $Z$ & $I$ \\
		\hline
		$O_2$ & $Z$  & $I$ & $Z$ & $Z$ & $I$ \\
		\hline
	\end{tabular}
\end{center}
The corresponding circuit $W$ can be easily constructed as a product of single-qubit Clifford gates.
This gives rise to a partition of $N$ qubits into five disjoint subsets,
$[N]=ABCDE$. For example, $A$ contains all qubits $j$ such that 
$O_1$ and $O_2$ act on $j$ by Pauli $X$ and $Z$ respectively. 
Note that $A$ has an odd size since otherwise $O_1$ and $O_2$ would commute.
In particular, $A\ne \emptyset$.

Suppose $B\ne \emptyset$. Apply $W=\mathsf{CZ}_{a,b}$ for some $a\in A$ and $b\in B$.
 This reduces $|O_1|$ by one without changing $O_2$. 
 
Suppose $CD\ne \emptyset$. Apply $W=\mathsf{CNOT}_{b,a}$ for some $a\in A$ and $b\in CD$. 
This reduces $|O_2|$ by one without changing $O_1$.
	
 In the remaining case $B=C=D=\emptyset$ and $A\ne \emptyset$. 
Assume wlog that $A=\{1,2,\ldots,2k+1\}$ for some integer $k$.
Then $O_1=X_1 X_2\cdots X_{2k+1}$ and $O_2=Z_1 Z_2 \cdots Z_{2k+1}$.
Set $q=1$. Choose
\[
W = \prod_{a=1}^k  \mathsf{CNOT}_{1,2a+1}\mathsf{CNOT}_{2a,1}\mathsf{CNOT}_{2a+1,2a}.
\]
One can easily check that $W^\dag O_1 W = X_1$ and $W^\dag O_2 W = Z_1$.
Thus $W$ reduces the combined weight $|O_1|+|O_2|$ by $4k$.
Furthermore, $W$ contains $3k\le 4k$ two-qubit gates.
\end{proof}
	
\begin{proof}[\bf Proof of Lemma~\ref{lemma:2}]
We shall  use the notations introduced in the  proof of Lemma~\ref{lemma:1}.
Consider the standard form of $O_1$ and $O_2$. 
Suppose $A\ne \emptyset$. Note that $|A|$ is even since $O_1$ and $O_2$ commute.
Assume wlog that $A=\{1,2,\ldots,2k\}$ for some integer $k$.
Choose
\[
W=\prod_{a=1}^k \mathsf{H}_{2a-1} \mathsf{CNOT}_{2a,2a-1} \mathsf{CNOT}_{2a-1,2a}.
\]
One can can easily check that $W$ maps $X_1 X_2 \ldots X_{2k}$
and $Z_1 Z_2 \ldots Z_{2k}$
to $Z_1 Z_3 \ldots Z_{2k-1}$ and $Z_2 Z_4 \cdots Z_{2k}$ respectively.
The combined weight $|O_1|+|O_2|$ is reduced by $2k$ and $W$
contains $2k$ two-qubit gates. 
From now on we can assume $A=\emptyset$.

If $B\ne \emptyset$ and $D\ne \emptyset$ then  apply $W=\mathsf{CNOT}_{d,b}$ for some $b\in B$ and $d\in D$.
This reduces $|O_1|$ by one without changing $O_2$.

If $C\ne \emptyset$ and $D\ne \emptyset$ then  apply $W=\mathsf{CNOT}_{d,c}$ for some $c\in C$ and $d\in D$.
This reduces $|O_2|$ by one without changing $O_1$.

After a sequence of steps as above we have $A,D=\emptyset$.
Note that $B\ne \emptyset$ and $C\ne \emptyset$ since we assumed 
that $O_1\ne I$ and $O_2\ne I$.

If $|B|\ge 2$ then apply $W=\mathsf{CNOT}_{b,b'}$ for some qubits $b,b'\in B$.
This reduces $|O_1|$ by one without changing $O_2$.
If $|C|\ge 2$ then apply $W=\mathsf{CNOT}_{c,c'}$ for some qubits $c,c'\in C$.
This reduces $|O_2|$ by one without changing $O_1$.

After a sequence of steps as above we have $A,D=\emptyset$,
$|B|=1$, and $|C|=1$, that is, $O_1 = Z_p$ and $O_2=Z_q$ for some
pair of qubits $p\ne q$.
\end{proof}

\subsection{State initialization routines} \label{stateprep}
For the entanglement-forging scheme used in the experiment, we need to initialize $N$ qubits in a superposition state $\ket{\phi_{xy}^p} = (\ket{x} + i^p\ket{y})/\sqrt{2}$, where $x$ and $y$ are $N$ qubit bitstrings, and $x \neq y$. A general construction of $\ket{\phi_{xy}^p}$ proceeds as follows:
\begin{enumerate}
\item Find an index $k$ where $x_k \neq y_k$. 
\item If $x_k=1$ ($y_k=0$), swap the definitions of $x$ and $y$, and substitute $p \rightarrow (-p\mod{4})$, using the fact that $\ket{\phi_{xy}^p}$ is the same as $\ket{\phi_{yx}^{-p} }$ up to an inconsequential global phase.
\item Find the sets of indices $S = \{ l \neq k : x_l \neq y_l \}$ and $T = \{ l : x_l = 1 \}$.
\item Given a set of $N$ qubits prepared in $|0\rangle^{\otimes N}$, apply the single-qubit gates $\otimes_{i \in T} \mathrm{X}_i$.
\item Apply single-qubit gate $G_p$ to qubit $k$, defined as $(G_0, G_1, G_2, G_3) = (\mathrm{H} , \mathrm{SH} , \mathrm{ZH} , \mathrm{SZH})$. Note the Hadamard acts first in each case.
\item \label{cnot_step} For all qubits $l \in S$, apply $\mathrm{CNOT}_{kl}$.
\end{enumerate}

The above procedure may become difficult to execute with high fidelity in simulations of large systems as limited processor connectivity may lead to large numbers of costly swap operations, or simply as the Hamming distance between pairs of included bitstrings grows and necessitates more CNOTs in step \ref{cnot_step}. Moreover, circuits with different initialization gates will experience different errors, and these uneven error rates can lead to inference of non-physical average behaviors, e.g. energies below the correct ground state energy. To circumvent this issue, one may replace the circuits initialized as $N$-qubit superposition states by a larger number of circuits initialized as products of single-qubit states, which are readily prepared with high fidelity. Suppose $x\ne y$ 
are $N$-bit strings that differ on $d$ bits. For each qubit $j$ and an integer $p$ define a single-qubit state
\begin{equation}
\label{single_qubit_prep1}
|\psi_{xy}^{pj}\ra = \left\{
\begin{array}{rcl}
|x_j\ra & \mbox{if} & x_j=y_j \\
\frac1{\sqrt{2}} \left( |x_j\ra + e^{i \pi  p/2d} |y_j\ra \right) &\mbox{if} & x_j\ne y_j \\
\end{array}
\right.
\end{equation}
Define an $N$-qubit tensor product state
\begin{equation}
\label{single_qubit_prep2}
|\psi_{xy}^p\ra = |\psi_{xy}^{p1}\ra \otimes |\psi_{xy}^{p2}\ra \otimes \cdots \otimes |\psi_{xy}^{pN}\ra.
\end{equation}
Such state can be easily prepared starting from the basis vector $|0^N\ra$ by applying
Hadamard gates, Pauli $X$ gates, and single-qubit $Z$-rotations by the angle $\pm \pi p/2d$.
After simple algebra one gets a decomposition
\begin{equation}
\label{product_forging1}
|x\ra\la y|^{\otimes 2} + |y\ra \la x|^{\otimes 2} 
=\frac{4^d}{4d} \, \sum_{p=0}^{4d-1} (-1)^p |\psi_{xy}^p \ra\la \psi_{xy}^p |^{\otimes 2}.
\end{equation}
Note that the righthand side is a linear combination of $2N$-qubit tensor product states. 
We can use this decomposition to classically forge entanglement for an arbitrary state
\[
|\psi\ra = (U\otimes V) \sum_n \lambda_n |b_n\ra \otimes |b_n\ra.
\]
Indeed, write the density matrix $|\psi\ra\la \psi|$ as a linear combination
of diagonal terms $|b_n\ra \la b_n|^{\otimes 2}$ and off-diagonal terms
$|b_n\ra\la b_m|^{\otimes 2} + |b_m\ra\la b_n|^{\otimes 2}$ with $n\ne m$.
Applying Eq.~(\ref{product_forging1}) with $x=b_n$ and $y=b_m$ to each off-diagonal term
one  finally arrives at
\begin{align}
\label{product_forging2}
\la \psi | O_1 \otimes O_2 |\psi\ra 
&=\sum_n \lambda_n^2  \la b_n | U^\dag O_1 U|b_n\ra \cdot  \la b_n | V^\dag O_2 V|b_n\ra \nonumber \\
& + \sum_{n<m}  \frac{\lambda_n \lambda_m 4^{d_{n,m}}}{4d_{n,m}} \,
\sum_{p=0}^{4d_{n,m}-1} (-1)^p
\la \psi_{b_n b_m}^p|U^\dag O_1 U| \psi_{b_nb_m}^p \ra \cdot
\la \psi_{b_n b_m}^p|V^\dag O_2 V| \psi_{b_nb_m}^p \ra.
\end{align}
where $O_i$ are arbitrary $N$-qubit observables and  
$d_{n,m}$ is the Hamming distance between the bit strings $b_n$ and $b_m$
(the number of bit flips separating $b_n$ and $b_m$).
Note that each term in Eq.~(\ref{product_forging2}) can be estimated on a device with only $N$ qubits 
by initializing each individual qubit in the state $|0\ra$, $|1\ra$, or $(|0\ra+e^{\pm i\pi p/2d}|1\ra)/\sqrt{2}$
according to Eqs.~(\ref{single_qubit_prep1},\ref{single_qubit_prep2})
with $x=b_n$, $y=b_m$, 
applying the circuit $U$ or $V$, and measuring the
eigenvalue of $O_1$ or $O_2$. 
However, the exponential factor $4^{d_{n,m}}$ in the  decomposition
Eq.~(\ref{product_forging2}) 
may lead to a loss of accuracy. For concreteness, suppose the observables
$O_i$ are normalized such that the operator norm of $O_i$ is at most one
(for example, $O_i$ are $N$-qubit Pauli operators). 
Suppose each $N$-qubit expected value
in Eq.~(\ref{product_forging2}) can be estimated within an additive error $\epsilon_0$.
Using the triangle inequality one can easily check that the righthand side
of Eq.~(\ref{product_forging2}) approximates the expected
value $\la \psi | O_1 \otimes O_2 |\psi\ra$ within an additive error 
\begin{equation}
\label{epsilon}
\epsilon =O(\epsilon_0) \left( 1 +  \sum_{n<m}  |\lambda_n \lambda_m| 4^{d_{n,m}} \right).
\end{equation}
Thus the method is practical only if $d_{n,m}$ is sufficiently small for all bit strings $b_n$ and $b_m$
that contribute to $|\psi\ra$.

\subsection{Connectivity and swaps}
\label{connectivity}
Figure 5 outlines how we utilized the structure of the particular forged ansatz in this demonstration to compile the required circuits onto a 5-qubit line without introducing any additional gate-based swap operations. Panel \textbf{a} shows the definition of the hop gate, denoted by green diamonds, which we may also abbreviate as a swap followed by the gate denoted by green circles. Panel \textbf{b} likewise defines a ``modified hop gate,'' which does not involve a swap, and acts like the hop gate except that it leaves the $\ket{11}$ state unchanged. Following panel \textbf{c}, we start from the set of hop gates in Fig. \ref{fig:ansatz}, use the fact that in the experiment we fix $\varphi_3 = 0$ to reduce that hop gate to a CPHASE gate, and unpack gates 1, 4, and 5 according to the hop gate definition. (The CPHASE gate is compiled using Hadamards and a CNOT).

\begin{figure}
    \centering
    \includegraphics[width=0.7\textwidth]{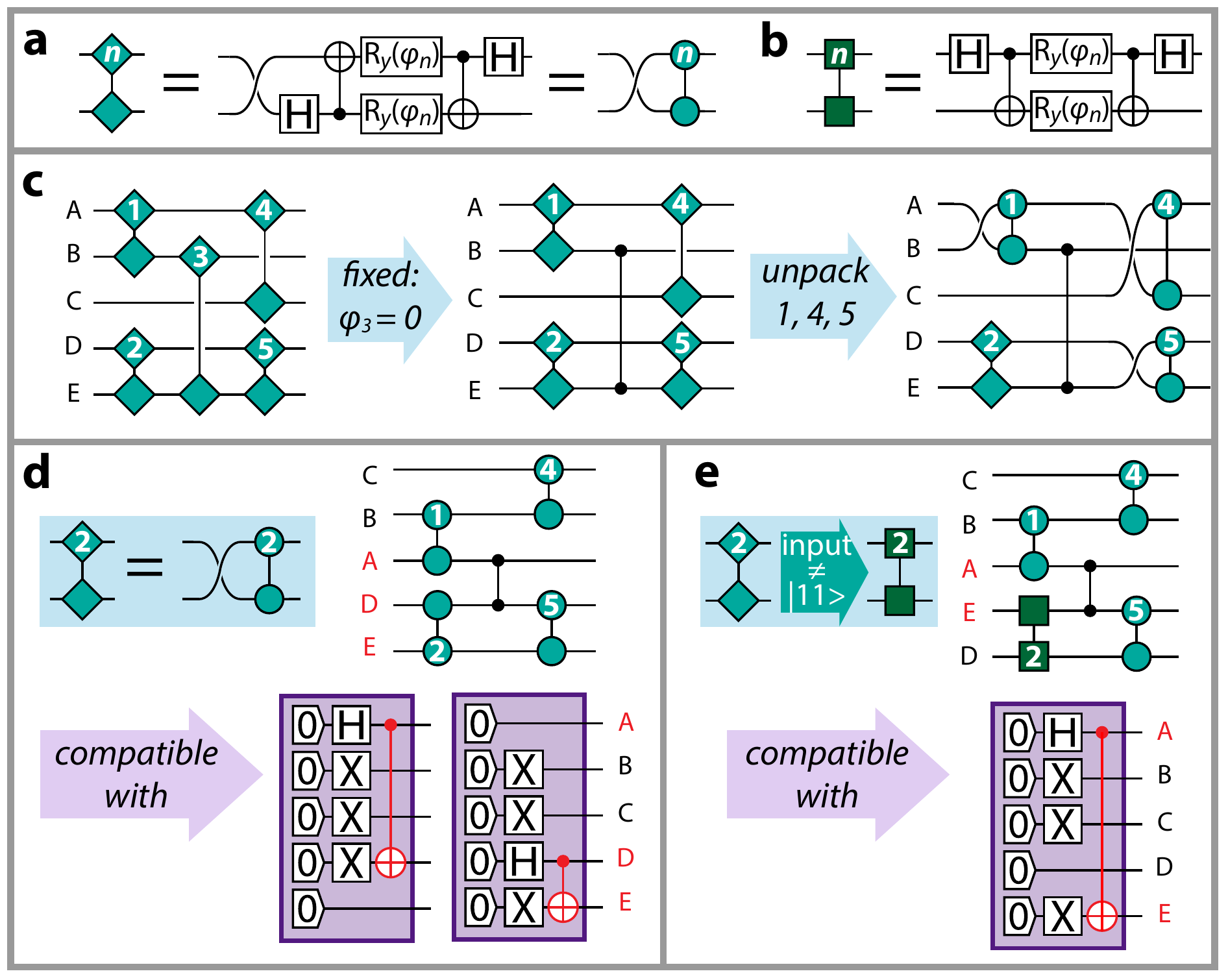}
    \label{fig:connectivity}
    \caption{\textbf{Compilation onto a line of qubits.} \textbf{a}, Definition of the hop gate (green diamonds) in terms of a swap followed by standard gate operations, the latter abbreviated by the gate with green circles. \textbf{b}, Definition of the modified hop gate, equivalent to a hop gate except that it leaves the $\ket{11}$ state unchanged. \textbf{c}, Initial compilations steps for the hop gates used in the experiment. \textbf{d}, Likewise unpacking hop gate 2 per its definition leads to the circuit in the upper right, the connectivity of which is compatible with the two state-initialization subcircuits shown (purple rectangles). \textbf{e} Alternatively, replacing hop-gate 2 with a modified hop gate provides a connectivity solution for the state-initialization subcircuit shown in this panel.}
\end{figure}

We now compile hop gate 2 differently depending on the initialization subcircuits (purple rectangles) of the given circuit. Unpacking gate 2 per its definition and untwisting all crossed wires produces the circuit in the top right of panel \textbf{d}. The wires have been permuted to reveal the linear connectivity; in terms of the letters labeling the qubits in the figure, the required connectivity is C-B-A-D-E. To this set of gates we will need to prepend initialization subcircuits preparing superpositions of bitstrings. Each such subcircuit in this experiment contains one additional CNOT (red); to avoid any gate-based swaps, the two involved qubits must also be connected. The two purple subcircuits in \textbf{d}, $\ket{\phi^0_{b_1 b_2}}$ and $\ket{\phi^0_{b_2 b_3}}$, respectively require connections A-D and D-E. Since this is compatible with the connectivity C-B-A-D-E, we compile hop gate 2 as per panel \textbf{d} whenever these states are prepared.

In contrast, superposing the third pair of bitstrings by the $\ket{\phi^0_{b_1 b_3}}$ subcircuit in \textbf{e} requires A-E connectivity. To realize this, we note that this state preparation ideally never provides $\ket{11}$ as an input to hop gate 2, and thus hop gate 2 may be replaced by a modified hop gate with no nominal change in behavior. With the swap removed from hop gate 2, the desired linear connectivity of the hop-gate circuit becomes C-B-A-E-D, providing the A-E connection facilitating $\ket{\phi^0_{b_1 b_3}}$.

Nonphysical results may be inferred in cases where hardware noise acts unevenly across the set of executed circuits, as will occur to some extent when varying gate compilations depending on the state preparation. However, the total error in the circuit is reduced by the avoidance of swaps, and we accept the tradeoff for this experiment. An intriguing question is how generally entanglement forging and other circuit decomposition techniques may be used to circumvent limitations of device connectivity \cite{mitarai2020constructing}. Future entanglement-forging experiments may reduce connectivity requirements using the decomposition to single-qubit stabilizer states presented in \ref{stateprep}.

\subsection{Weighted sampling of circuits: Theory}
\label{weightedSamplingTheory}
Suppose $O_1,O_2 \in \{I,X,Y,Z\}^{\otimes N}$ are $N$-qubit Pauli observables 
and $|\psi\ra$ is a $2N$-qubit state.
Entanglement forging provides a decomposition
\be
\label{MCeq1}
\la \psi|O_1\otimes O_2|\psi\ra = \sum_{a=1}^\ell \mu_a \mathrm{Tr}(O_1 \rho_a)  \mathrm{Tr}(O_2 \rho_a),
\ee
where $\rho_a$ are  $N$-qubit pure states
and $\mu_a$ are real coefficients simply related to the Schmidt coefficients of $|\psi\ra$.
Let $V_i$ be a Clifford circuit  that maps each Pauli term $X$ or $Y$ that appears in $O_i$
to the Pauli $Z$.  Note that $V_i$ includes only single-qubit Clifford gates. Then
\[
O_i = V_i^\dag \left( \sum_{x\in \{0,1\}^N} O_i(x) |x\ra\la x|\right) V_i
\]
where $O_i(x)$ takes value $+1$ or $-1$ if the parity of $x$
restricted  to the support of $O_i$ is even or odd respectively.
We would like to estimate the quantity $\la \psi|O_1\otimes O_2|\psi\ra$ on a quantum device
with only $N$ qubits by a series of experiments such that each experiment
prepares a state $V_i \rho_a V_i^\dag$
for some pair $(i,a)$ and measures each qubit
in the computational basis. 
We claim that estimating $\la \psi|O_1\otimes O_2|\psi\ra$ with an additive error
$\epsilon$ and a confidence level $99\%$ requires at most $S$ experiments, where
\begin{equation}
\label{MCeq2}
S = \frac{200\left( \|\mu\|_1 \right)^2}{\epsilon^2},
\qquad  \|\mu\|_1 \equiv \sum_{a=1}^\ell |\mu_a|.
\end{equation}
For large problems, Monte Carlo methods may be used to generate the necessary $M=S/2$ state preparations $\rho_a$. Indeed, define a normalized probability distribution
$\pi_a = |\mu_a|/\|\mu\|_1$, where  $a=1,2,\ldots,\ell$, and let $a_1,a_2,\ldots,a_M\in \{1,2,\ldots,\ell\}$ be $M$ independent samples 
from the distribution.
For each $j=1,2,\ldots,M$
perform two experiments: (1) prepare the state $V_1 \rho_{a_j} V_1^\dag$,
measure every qubit in the $Z$-basis, and record the measured
bit string
$x^j  \in \{0,1\}^N$; (2) prepare the state $V_2 \rho_{a_j} V_2^\dag$,
measure every qubit in the $Z$-basis, and record the measured
bit string $y^j  \in \{0,1\}^N$. Define a random variable
\be
\label{MCeq3}
f = \frac{\|\mu\|_1}M \sum_{j=1}^M \mathrm{sgn}(\mu_{a_j}) O_1(x^j) O_2(y^j).
\ee
One can easily check that the mean and the variance of $f$ are 
\begin{equation}
\label{MCeq4}
\mathbb{E}(f) = \la \psi|O_1\otimes O_2|\psi\ra \quad \mbox{and} \quad
\mathbb{E}(f^2) -\mathbb{E}(f)^2=\frac{\|\mu\|_1^2 - \la \psi|O_1\otimes O_2|\psi\ra^2}{M}.
\end{equation}
Here the expectation values are taken over the random choice of $a_1,\ldots,a_M$ 
and the random measurement outcomes.
We conclude that $f$ is an unbiased estimator of $\la \psi|O_1\otimes O_2|\psi\ra$ with the variance 
at most $\|\mu\|_1^2/M=\epsilon^2/100$.
By the Chebyshev inequality, 
$|f-\la \psi|O_1\otimes O_2|\psi\ra| \le \epsilon$ with probability at least $0.99$.

We note that estimating $\la \psi|O_1\otimes O_2|\psi\ra$ on a $2N$-qubit device 
with a precision $\epsilon$ and confidence level $99\%$ 
would require at most $100\epsilon^{-2}$ experiments
(each experiment prepares the state $V_1\otimes V_2|\psi\ra$ and measures every qubit
in the $Z$-basis).
 From Eq.~(\ref{MCeq2})
one infers that entanglement forging increases the required number of experiments
roughly by the factor $2(\|\mu\|_1)^2$. However, each experiment requires $N$ instead of $2N$ qubits.
In addition, 
as argued in the main text, quantum circuits preparing the states $\rho_a$ are much simpler
compared with the circuit preparing the full state $\psi$.

Let  $\lambda=(\lambda_1,\lambda_2,\ldots,\lambda_{2^N})$ be the Schmidt coefficients of $|\psi\ra$.
Eq.~(\ref{densityOp}) from the main text gives
\be
\label{MCeq5}
\|\mu\|_1 =  1 + 4\left( \sum_{i=1}^{2^N} |\lambda_i| \right)^2.
\ee
Thus entanglement forging is mostly useful for weakly entangled states 
such that  $\|\lambda\|_1 = \sum_{i=1}^{2^N} |\lambda_i|$ is a constant or a slowly growing function of $N$
(note that $\|\lambda\|_1 = \sqrt{2^N}$ in the worst case when $|\psi\ra$ is a maximally entangled
state of $N+N$ qubits).
In practice, one may wish to limit the maximum number of experiments $S$  by some specified cutoff $S_{max}$. 
Combining  Eqs.~(\ref{MCeq2},\ref{MCeq5})  that determine the number of experiments
$S=S(\epsilon,\lambda)$  one obtains a non-linear constraint 
$S(\epsilon,\lambda) \le S_{max}$ on the vector of Schmidt coefficients $\lambda$.
This constraint should be incorporated into the classical optimizer that minimizes
the variational energy. 

\subsection{Weighted sampling of circuits: Implementation}
\label{weightedSamplingExpt}
As the number of bitstrings included in the experimental demonstration (3 bitstrings) was computationally tractable, Monte Carlo methods were not required. To minimize the variance of our estimate of the sum in equation \ref{expectation}, we wish to allocate the total experimental samples $S$ to the various circuits $\braket{b_n|\tO_k|b_m}$ in proportion to the respective coefficients $\lambda_n\lambda_m$. However, at the time of the experiment Qiskit did not yet support non-uniform sampling of a list of distinct quantum circuits, instead sampling all $J$ circuits in the submitted list, or job, an equal number of times $s = S/J$. We thus wrote a small ``copysampling'' function to approximate non-uniform sampling by populating the job with proportionally more copies of more important circuits. The function takes as inputs the desired normalized statistical weight $w_c$ for each circuit $c$, and the final size $J$ of the job to be submitted. Thus a target expression for the number of copies desired for circuit $c$ is $w_c J$, though this is not necessarily integer-valued. For simplicity, the copysampling function first ensures the job includes at least one copy of every possible circuit. Next, it deterministically appends $\mathrm{floor}(w_c J -1)$ additional copies of each circuit $c$. There are then $J - \sum_c\big( 1+\mathrm{floor}(w_c J -1)\big)$ remaining spots in the job. These are allocated to circuits randomly according to the remaining residuals, without replacement, to approximate the desired weighting distribution. A similar partially-deterministic shot allocation scheme was studied in \cite{arrasmith2020operator}. The job is then executed, and the results (``counts'') for all copies of $c$ are merged before analysis.

The above is limited in that the weighted-sampling is coarse-grained by the number of samples $s$ per copy of each circuit. The desired weighted-sampling distribution could be better approximated by setting $s=1$ and $J=S$. However, at present there is appreciable overhead in execution time scaling with the job size $J$. In our experiment, we found $J\approx 800$ and $s\approx 2000$ gave an acceptable compromise between job-size overhead time and efficiency of weighted sampling.

Our sampling weights accounted for both the coefficients $\lambda_n\lambda_m$ and the zero-noise extrapolation stretch factors discussed below. We note that possible extensions could further optimize sampling weights for each quantum circuit by accounting for weights of the Pauli strings in the Hamiltonian, optionally further informed by the expectations of Pauli strings obtained in the previous VQE iteration. Optimally assigning samples accounting for Pauli strings is nontrivial as Hamiltonian terms contain pairwise products of $N$-qubit Pauli-string expectations, and moreover groups of compatible Paulis are in our experiment measured jointly via the tensor-product basis grouping method \cite{Kandala2017} standard in Qiskit. Nonetheless, preliminary simulations suggest such weighting schemes can further reduce the required number of samples $S$ by more than $\sim25\%$.

\subsection{Repeated-gate zero-noise extrapolation}
\label{ZNE}
The accuracy of energies computed on the quantum hardware was significantly improved by mitigating gate errors via zero-noise extrapolation, or ZNE \cite{Kandala2019,richardsonTheory}. To avoid the overhead of calibrating stretched gates, we implemented a repeated-gate ZNE routine similar to that in \cite{mitiq}. For each quantum circuit, a copy is made by applying the transformation $G\rightarrow GG^{-1}G$ to each primitive gate, approximating a noise-amplification factor of 3. We use first-order extrapolation of the results from the original and copied circuits to infer the approximate zero-noise result for each distinct circuit (i.e. for each distinct [state-preparation, Pauli-string] combination) used to compute the desired observable. Additionally, because the extrapolated result depends more sensitively on the original circuit than on the noise-amplified circuit, to improve overall precision we assign a proportionally higher weight to the original circuit in the weighted sampling routine described above.

\subsection{Parallel circuit execution}
\label{parallelization}
To reduce the total data acquisition time, we executed pairs of VQE problems simultaneously on different subsets of the {\tt Dublin} processor. Physical qubits were selected based on automated device benchmarking performed shortly before to the start of each VQE run. To reduce the risk of crosstalk between problems, we required that no direct connectivity exist between the two sets of physical qubits, with a buffer of at least one idle physical qubit. A similar parallelization scheme was recently characterized in \cite{niu2021enabling}. Future parallelized experiments might suppress readout crosstalk as needed by choosing sets of qubits disjoint with respect to readout-line multiplexing, or by applying efficient mitigation methods such as in \cite{bravyi2020mitigating}.

This parallelization routine was implemented in software to be a modular wrapper of the experiment code. Using the standard Python {\tt multiprocessing} package, the wrapper first launches a Python process executing one copy of the experiment code for each independent VQE problem. In each iteration of the VQE optimization, each of these processes generates a list of circuits, or job, it needs executed. Rather than passing the job directly to the backend as usual, each process passes its job to a common ``multiplexer'' Python process. The multiplexer respectively merges the circuits in the separate jobs into a single new job, sends that to the backend for execution, and then separates the component results and returns them to the corresponding VQE processes. This framework is convenient as the multiplexer can be swapped in as a virtual backend with little modification of the original experiment code.

\subsection{Schmidt coefficients and truncation}
\label{svdTruncation}
\begin{figure}
    \centering
    \includegraphics[width=0.85\textwidth]{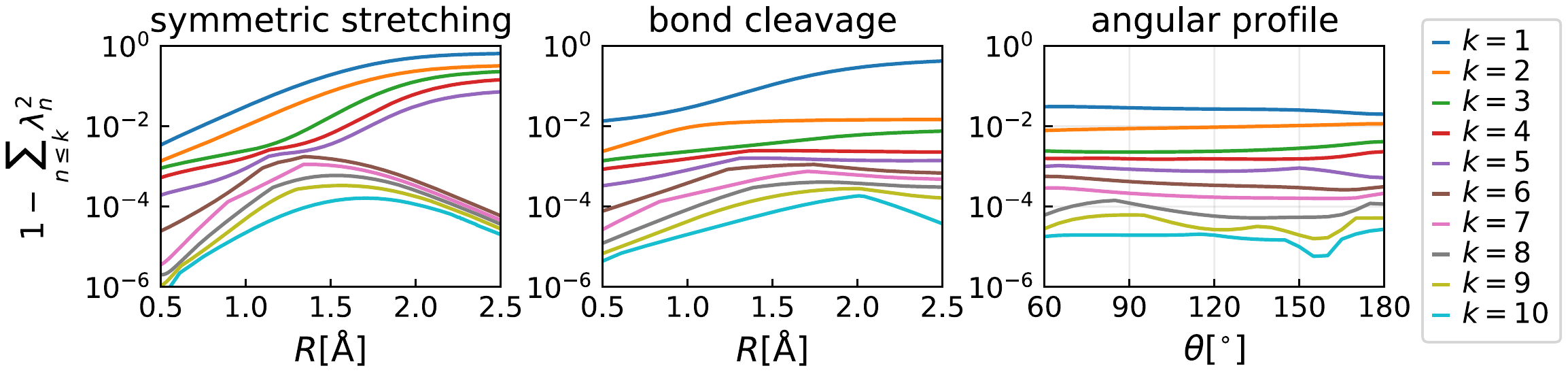}
    \label{fig:SVD}
    \caption{\textbf{Accuracy of truncated Schmidt decompositions.} The three panels correspond to the three sweeps of molecular geometries as in Figs. \ref{fig:energy_plots},\ref{fig:singular_values}. For each choice of bitstring truncation $k$, we plot one minus the projection of the FCI wavefunction into the space defined by the leading $k$ bitstrings. In other words, we plot the sum of squares of the FCI Schmidt coefficients excluded by the truncation. Values near 1 (0) indicate that the FCI wavefunction cannot (can) be accurately represented in the truncated space. Unlike in the rest of the text, these FCI calculations do not freeze the oxygen $2p$ orbital.}
\end{figure}

For technical convenience in the forging demonstration, rather than sampling all possible bitstrings, we truncated the space to include only the leading $k=3$ bitstrings, that is, set $\lambda_{n>k}=0$. Noiseless simulations with $k=3$ and $k=6$ were performed for comparison (Fig. \ref{fig:energy_plots}). Figure \ref{fig:SVD} illustrates how the choice of $k$ limits accuracy of describing the various ground state wavefunctions, here obtained via FCI calculations. For the case of symmetric stretching, a qualitative change occurs for $k\geq6$ such that accuracy improves as the molecule dissociates further. Single-bond cleavage exhibits a similar, though less pronounced, change for $k\geq2$, as per the discussion in the main text. In contrast, sweeping the bond angle does not reveal any critical value of $k$. Note that unlike the experiments in the main text, the FCI calculations here do not freeze the oxygen $2p$ orbital; including this orbital increases the number of allowed bitstrings from 10 to 15, such that a nonzero residual remains for $k=10$.

\subsection{Updating Schmidt coefficients}
\label{updatingLambda}
The Schmidt coefficients $\lambda_n$ must be updated each iteration of the VQE optimization. One way to do this is by treating them on the same footing as the gate parameters ($\varphi$ in Fig. \ref{fig:ansatz}), such that the optimizer explores a space with dimension equal to the number of Schmidt coefficients plus the number of gate parameters. Here however, since the problems are small, we instead can and do remove the Schmidt coefficients from the optimizer search space by having the classical processor exactly minimize the energy as a function of the Schmidt coefficients based on the latest measurement results. This process is detailed below.

In each iteration of the VQE, quantum circuits are sampled to estimate the expectation values $\braket{b_n|\tO_i|b_n}$ and $\braket{\phi^p_{b_{n}b_{m}}|\tO_i|\phi^p_{b_{n}b_{m}}}$ in Eq. \ref{expectation}, where $O_i$ here represents an $N$-qubit Pauli string. Computing the energy involves plugging these values into a product over $i$ (indicating spin-up or -down), and summations over $p, n, m$, and the weighted list of $2N$-qubit Pauli strings defining the Hamiltonian. Deferring the operations over Schmidt-coefficient indices $n, m$ and performing all others leaves
\begin{equation}
    \langle H \rangle = \sum_{n,m} \lambda_n\lambda_m h_{nm},
\end{equation}
where $h$ is a symmetric matrix of known values. By the variational principle, the eigenvector of this matrix with the lowest eigenvalue minimizes the energy. Thus we update $\lambda$ by computing this eigenvector, such that the VQE optimizer need only minimize $E(\varphi)$ rather than $E(\varphi, \lambda)$.

Although Schmidt coefficients are typically taken to be non-negative for definiteness, in this computation we allow them to take any real values, facilitating the exact-minimization update routine above and increasing ansatz flexibility for fixed $U,V$. In the main text, empirical $\lambda$ values (Fig. \ref{fig:singular_values}) are represented by their absolute values for simplicity of discussion.

\subsection{Example classical simulation with a larger ansatz}
\label{largerAnsatz}
We executed an additional classical simulation with a larger ansatz at the nominal equilibrium geometry (defined in main text) to emphasize that the entanglement forging decomposition is exact, and thus with a sufficiently large ansatz and low errors can obtain arbitrary accuracy. Moreover, the $U$ (and $V$) part of the ansatz can always be constructed exclusively from hop gates. We targeted a discrepancy of 1.6 mH or less from the active-space FCI, as this value is sometimes quoted as a standard for quantum chemistry calculations. The ansatz included all 10 bitstrings consistent with assigning 3 electron-pairs to the 5 molecular orbitals in the active space ($k=10$), and the gates were found via a heuristic empirical search using a noiseless classical simulator. The optimization converged to a ground-state energy of -75.726303 H, leaving a discrepancy of 1.47 mH relative to the FCI value in this active space, -75.727775 H.

Explicitly, the $N$-qubit bitstrings, including frozen orbitals, are
\begin{center}
\begin{tabular}{ c c c c c }
 1111100 & 1011101 & 1011110 & 1101110 & 1101101 \\ 
 1110110 & 1110101 & 1001111 & 1010111 & 1100111
\end{tabular}.
\end{center}
The hop gates, listed in terms of the ordered orbital pair they acted upon (indexing from 0 to 6, such that here 0 and 4 are frozen), are given in the table below. This ad hoc sequence of gates is presumably not optimal in terms of circuit depth.
\setlength{\columnsep}{1cm}
\begin{center}
\begin{tabular}{r | c c}
& orbitals & $\theta$ (rad)\\ \hline
1 & 1,2 & 1.57107008e+00\\
2 & 5,6 & 7.85631357e-01\\
3 & 2,6 & 0 \\
4 & 1,3 & -1.64124047e-01\\
5 & 5,6 & 6.94946136e-01\\
6 & 2,3 & 0 \\
7 & 3,5 & -1.32698309e-03\\
8 & 3,5 & 0 \\
9 & 1,3 & 7.47539070e-02\\
10 & 2,6 & 2.73733721e-04\\
11 & 2,6 & 0 \\
12 & 1,2 & 8.04994286e-01\\
13 & 5,6 & 8.85249894e-01\\
14 & 2,6 & 0 \\
15 & 1,3 & -1.01079874e+00\\
16 & 5,6 & 7.98610796e-01\\
17 & 2,3 & 0 \\
18 & 3,5 & 1.79900833e-03\\
19 & 3,5 & 0 \\
20 & 1,3 & -6.89839195e-02\\
21 & 2,6 & 2.45242043e-03\\
22 & 2,6 & 0 \\
23 & 1,2 & 1.20565841e-03\\
24 & 5,6 & -1.23329417e-02\\
25 & 2,6 & 0 \\
26 & 1,3 & 1.68445871e-02\\
27 & 5,6 & -9.59290116e-03\\
28 & 2,3 & 0 \\
29 & 3,5 & -1.11378980e-02\\
30 & 3,5 & 0 \\
31 & 1,3 & 7.22490009e-03\\
32 & 2,6 & -8.25656996e-04\\
33 & 2,6 & 0
\end{tabular}.
\end{center}
\clearpage

\end{document}